\documentclass[final]{aipproc}
\usepackage{latexsym}
\usepackage{epsfig}
\layoutstyle{6x9}

\begin{document}

\title{Double Higgs Production at the LHC as a Robust Test of
Little Higgs Models}


\author{Claudio O. Dib}
{address={Dept.\, of Physics, Universidad T\'ecnica Federico Santa
Mar\'{\i}a, Valpara\'{\i}so, Chile} }

\author{Rogerio Rosenfeld}
{address={Instituto de F\'{\i}sica Te\'orica - UNESP,
 Rua Pamplona, 145, 01405-900 S\~ao Paulo, SP, Brasil} }

\author{Alfonso Zerwekh}
{address={Department of Physics, Universidad Austral de Chile,
 Valdivia, Chile} }

\begin{abstract}
We analyze double Higgs boson production at the LHC in the context
of Little Higgs models. In double Higgs production, the diagrams
involved are directly related to those that cause the cancellation
of the quadratic divergence of the Higgs self-energy, so this mode
provides a robust prediction for this class of models. We find
that in extensions of this model with the inclusion of a so-called
T-parity, there is a significant enhancement in the cross sections
as compared to the Standard Model.
\end{abstract}

\maketitle

The presence of quadratic divergences in loop corrections to the
scalar Higgs boson self-energy is responsible for the so-called
hierarchy problem of the Standard Model (SM); namely, there is no
natural way of having a ``light'' mass (i.e. $\sim 10^2$ GeV) for
the Higgs given that loop corrections induce mass terms of the
order of the scale at which new physics enters-- be it the GUT
scale or any other above a few TeV. In Supersymmetric extensions
this problem is absent since the divergence in bosons and fermions
are related and the latter can only be logarithmic.\cite{susy}. It
is also absent in models where scalar particles are not
fundamental but composite\cite{techni}.

Recently a new kind, called Little Higgs (LH) model \cite{little},
which can solve the hierarchy problem was proposed. Here the Higgs
is a pseudo-Goldstone boson whose mass is protected by a global
symmetry and, unlike supersymmetry, quadratic divergence
cancellations are due to contributions from new particles with the
same spin.

The phenomenology of these models has been discussed with respect
to indirect effects on precision measurements \cite{precision} and
direct production of the new particles introduced \cite{pheno}.
Since these early contributions, several variations have been
proposed \cite{review}, but the cancellation of quadratic
divergences is inherent to any LH model and this requires definite
relations among certain couplings. Therefore, any process that
involves exclusively these couplings is a robust prediction of the
LH mechanism regardless of model variations. One of such processes
is double Higgs production, which we study here.

After the spontaneous breakdown of a global underlying symmetry at
a scale $4\pi f$ of a few TeV, the model contains a large
multiplet of pseudo-Goldstone bosons, which includes the SM Higgs
doublet. While most members of the multiplet receive large masses
(of a few TeV), the mass of the Higgs boson is protected from
quadratic divergences at one loop, and therefore remains naturally
smaller. The cancellation is related to the existence of an extra
(heavier) top-like quark and its interactions with the scalar
sector, feature which is common to all Little Higgs models. Higgs
pair production at LHC is based on exactly the same diagrams that
enter the quadratic divergence cancellation (Fig.~1), except for
the insertion of two gluons (Fig.~2). In order to work out the
details, we make use of the Littlest Higgs model, which is a
simple case but contains all the necessary features.

Below the scale $4\pi f$, the Little Higgs lagrangian
\cite{littlest} can be written as a non-linear sigma model based
on a coset $SU(5)/SO(5)$ symmetry:
\begin{equation}
{ \cal L}_{\Sigma} = \frac{1}{2} \frac{f^2}{4}
        {\rm Tr} | {\cal D}_{\mu} \Sigma |^2,
\label{Sigma}
\end{equation}
where the subgroup $[SU(2)\times U(1)]^2$ of $SU(5)$ is promoted
to a local gauge symmetry. The covariant derivative is defined as
${\cal D}_\mu \Sigma=  \partial_\mu\Sigma - i \sum_{j=1}^2\Big(
g_j( W_j\Sigma +  \Sigma W_j^T) + g'_j (B_j\Sigma + \Sigma B_j^T)
\Big)$. To exhibit the interactions, one can expand $\Sigma$  in
powers of ${1}/{f}$ around its vacuum expectation value $\Sigma_0$
\begin{equation}
\Sigma = \Sigma_0 + \frac{2 i}{f} \left(
\begin{array}{ccccc}
\phi^{\dagger} & \frac{h^{\dagger}}{\sqrt{2}} &
{\mathbf{0}}_{2\times 2} \\
\frac{h^{*}}{\sqrt{2}} & 0 &
\frac{h}{\sqrt{2}} \\
{\mathbf{0}}_{2\times 2} & \frac{h^{T}}{\sqrt{2}} &
\phi
\end{array} \right)
+ {\cal O}\left(\frac{1}{f^2}\right),
\end{equation}
where $h$ is the doublet that will remain light and $\phi$ is a
triplet under the unbroken $SU(2)$. The non-zero v.e.v. $\Sigma_0$
of the field leads to the breaking of global $SU(5)$ into $SO(5)$
and also breaks the gauge symmetry $[SU(2)\times U(1)]^2$ into its
diagonal subgroup, which is identified with the standard model
$SU_L(2)\times U_Y(1)$ symmetry group.

The standard model fermions acquire their masses via the usual
Yukawa interactions. However, in order to cancel  the top quark
quadratic contribution to the Higgs self-energy, a new vector-like
color triplet fermion pair, $\tilde t$ and $\tilde t^{\prime c}$,
with quantum numbers $({\mathbf{3,1}})_{Y_i}$ and
$({\mathbf{\bar{3},1}})_{-Y_i}$ must be introduced. Since they are
vector-like, they are allowed to have a bare mass term which is
{\it chosen} such as to cancel the quadratic divergence above
scale $f$. Accordingly, the standard top quark couples to the
pseudo-Goldstone bosons and heavy colored fermions in the littlest
Higgs model as:
\begin{equation}
{\cal{L}}_Y = \frac{1}{2}\lambda_1 f \epsilon_{ijk} \epsilon_{xy}
\chi_i \Sigma_{jx} \Sigma_{ky} u^{\prime c}_3 + \lambda_2 f
\tilde{t} \tilde{t}^{\prime c} + {\rm h.c.}, \label{yuk}
\end{equation}
where $\chi_i=(b_3, t_3, \tilde{t})$, $\epsilon_{ijk}$ and
$\epsilon_{xy}$ are antisymmetric tensors, and $\lambda_1,\
\lambda_2$ are parameters of order unity.

\begin{figure}
\epsfig{file=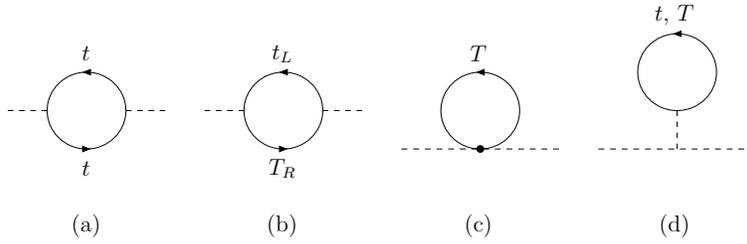,height=3.5cm,width=10cm}%
\caption{One-loop corrections to the Higgs mass, to order $v/f$:
(a) standard top quark loop, (b) mixture of standard and extra top
quark loop, (c) extra top quark loop with a 4-particle vertex, and
(d) tadpoles with standard and with extra top quark loops. There
are other diagrams but they are suppressed by factors of order
$(v/f)^2$ or higher.} \label{quadratic}
\end{figure}

After EWSB, we write $h^0 = 1/\sqrt{2} (v + H)$, and follow
Perelstein {\it et al.} \cite{PPP} in defining left handed fields
$t_{3 L} \equiv t_3$, $\tilde{t}_L \equiv \tilde{t}$ and right
handed fields $\bar{u}_{3 R}^{\prime}\equiv u_3^{\prime c}$,
$\bar{\tilde{t}}^{\prime}_{R} \equiv \tilde{t}^{\prime c}$ to
obtain
\begin{eqnarray}
{\cal L}_{t} &=&
- \left( \begin{array}{cc} \bar{u}_{3 R}^{\prime} & \bar{\tilde{t}}^{\prime}_{R}
\end{array} \right)
\left( \begin{array}{cc}  \lambda_1 v & \lambda_1 f (1-v^2/f^2) \\
0 & \lambda_2 f
\end{array} \right)
\left( \begin{array}{c}  t_{3 L} \\ \tilde{t}_L
\end{array} \right)
- \lambda_1 H \bar{u}_{3 R}^{\prime}  t_{3 L} + \\ \nonumber
 && \lambda_1 \frac{v}{f} H \bar{u}_{3 R}^{\prime} \tilde{t}_L +
 \frac{\lambda_1}{2 f} H^2 \bar{u}_{3 R}^{\prime} \tilde{t}_L +
h.c.
\end{eqnarray}
Diagonalizing this mass matrix, we obtain the usual eigenvalues
corresponding to the top quark $t$ and the heavy top $T$ masses,
$m_t$ and $m_T$, in terms of the scales $v$ and $f$, and the
couplings $\lambda_i$. From this analysis one also derives the
couplings of the Higgs to the top quarks $t_{L,R}$ and $T_{L,R}$
(of left and right chirality), in terms of the same parameters. In
an obvious notation, these couplings are denoted as $g_{Htt}$,
$g_{HT_Rt_L}$, $g_{Ht_RT_L}$, $g_{HTT}$, $g_{HHTT}$ and
$g_{HHtt}$. One should be aware that, for real values of
$\lambda_i$, the values of $m_t$, $m_T$, $v$ and $f$ not only are
related but also restricted to the condition:~\cite{DRZ}

\begin{equation}
m_T > 2 \frac{m_t{}v} f \simeq \sqrt{2} f .
\end{equation}
The relevant Feynman diagrams for the Higgs self-energy are shown
in Fig.~\ref{quadratic}.

%
%
%
%-----------FIGURE 1 --------------------------------------

The cancellation of tadpole diagrams requires that $g_{Htt} m_t +
g_{HTT} m_T = 0$, whereas the cancellation of higgs self-energy
quadratic divergences implies $g_{Htt}^2 + g_{HTT}^2 +
g_{HT_Rt_L}^2 + g_{Ht_RT_L}^2 + g_{HHtt} m_t + g_{HHTT} m_T = 0$.
These conditions are satisfied up to terms of order ${\cal O}
(v/f)$ by the masses and  couplings defined above.

An important point to consider is that in the simplest LH models,
strict bounds on the parameters exist. In particular, electroweak
precision constraints require $f > 3.5$ TeV \cite{precision}.
However, in a recent variation on the littlest Higgs model, where
a so-called T-parity that interchanges the two subgroups
$[SU(2)\times U(1)]_1$ and $[SU(2)\times U(1)]_2$ of $SU(5)$ is
introduced, this bound can be significantly lowered to $f > 500$
GeV \cite{TParity}. In this model, the T-odd states do not
participate in the cancellation of quadratic divergences and in
this respect our calculation is valid in this model as well.
T-parity also forbids the generation of a vacuum expectation value
for the triplet scalar field (i.e., $v'=0$ in the notation of
T.~Han {\it et al.}\cite{pheno}), which is one of the causes for
easing the electroweak constraints.

\begin{figure}
\epsfig{file=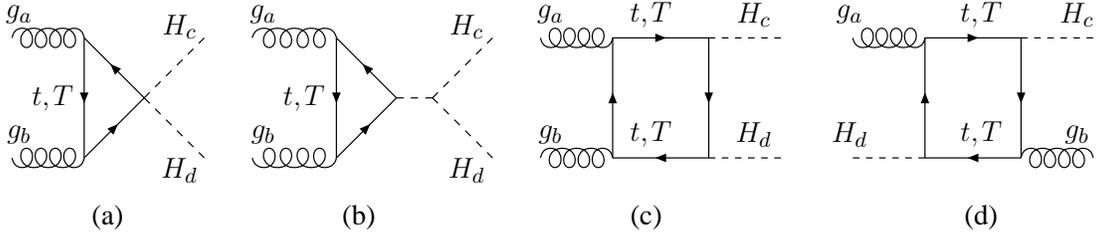,width=14.5cm}%
\caption{Contributions to Higgs boson pair production at LHC in a
Little Higgs model: (a) and (b) triangles; (c) planar boxes; (d)
non-planar boxes.} \label{triangles}
\end{figure}

Gluon-gluon fusion is the dominant mechanism of SM Higgs boson
pair production at the LHC \cite{dhiggs}. The amplitude for
 $gg \rightarrow HH$ process has contributions from triangle
and box diagrams~\cite{DRZ}, shown in Fig.~\ref{triangles}.
%
%
%
%
%-----------FIGURE 2 --------------------------------------
All these diagrams involve integrals that can be converted to
Passarino-Veltmann functions, for which there are computer codes
to evaluate them. Here the expressions for the amplitudes in terms
of Passarino-Veltman functions were computed using the package
FeynCalc \cite{FeynCalc} and the numerical integration of these
functions is done using LoopTools \cite{looptools}.

From the sum of all these diagrams and squaring, the partonic
differential cross section is obtained (we have included a factor
of $1/2$ from identical particles in the final state)
\begin{equation}
\frac{d \hat{\sigma}}{d \Omega} = \frac{1}{128 \pi^2 \hat{s}}
\sqrt{1 - 4 M_H^2/\hat{s}} \;\overline{| {\cal M}|^2}
\end{equation}
where $\overline{| {\cal M}|}^2$ is averaged over all 32 initial
color and helicity states. The $pp \rightarrow HH$ cross section
at LHC is then obtained by convoluting this partonic cross section
with the gluon distribution functions:
\begin{eqnarray}
\sigma (pp \to HH) &=& \frac{1}{2}   \int dx_1 dx_2 \;
[g_1(x_1,Q^2) g_2(x_2,Q^2) + g_2(x_1,Q^2) g_1(x_2,Q^2)]
\nonumber\\
 &&\qquad\qquad\hat{\sigma} (gg \to HH) \theta (x_1 x_2 s - 4
M_H^2).
\end{eqnarray}
Here we used the Set 3 of CTEQ6 Leading Gluon Distribution
Function with momentum scale $Q^2 = \hat{s}$ \cite{CTEQ}. A $K=2$
factor was included to take into account QCD corrections.

\begin{figure}
\epsfig{file=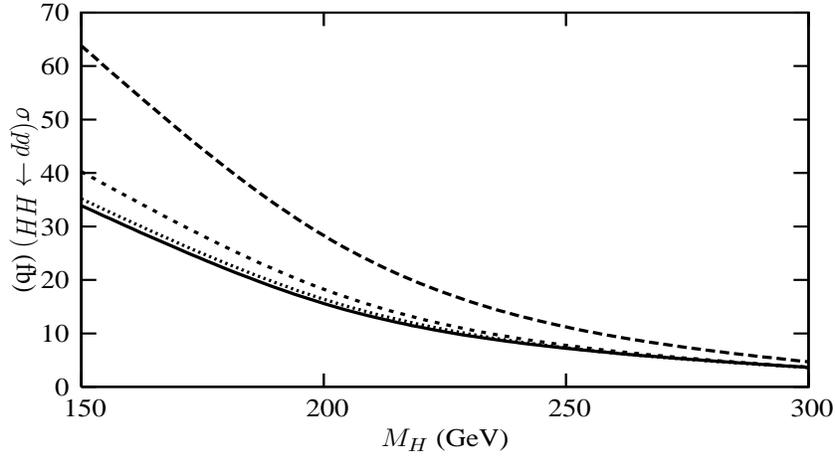,height=6cm, width=11cm} %
\caption{Cross section for double Higgs production at the LHC for
$M_T = 4$ TeV and $f =500$ GeV (dashed line), $1000$ GeV (short
dashed line) and $2000$ GeV (dotted line). In solid line is shown
the SM result.}\label{result1}
\end{figure}

In Fig.~\ref{result1} we plot the cross section for the double
Higgs production process at the LHC for fixed $M_T = 4$ TeV, a
Higgs boson mass in the range $150$--$300$ GeV and for $f =500$,
$1000$ and $2000$ GeV. As expected, we find that the largest
deviations from the SM result occurs for small Higgs boson mass
and small decay constant $f$.

%
%
%
%%-----------FIGURE 3 --------------------------------------

We also explored the dependence of this cross section on the mass
of the heavy top quark and found that it slowly grows with $m_T$,
but promptly reaches an asymptotic value, becoming insensitive for
$m_T > 2.5$ TeV.

%\begin{center}
\section*{Conclusions}
%\end{center}

Double Higgs production distinguishes Little Higgs (LH) models
from other electroweak symmetry breaking scenarios. The process is
intimately tied to the cancellation of quadratic divergences in
these models. We studied the reach of the LHC to probe the LH
models in this way. We found that only for relatively small values
of the energy scale $f$, of the order of $500$ to $1000$ GeV, it
is possible to distinguish meaningfully the LH from the SM. These
low values are attainable without violating electroweak precision
limits only in models where an extra T parity is incorporated
\cite{TParity}. These results are only mildly dependent on the
heavy top quark mass $m_T$.

\section*{Acknowledgements}

C.D. and R.R. thank the organizing committee of this conference
for the warm hospitality. A.Z. and C.D. received partial support
from Fondecyt (Chile) grants No.~3020002, 1030254, 7030107 and
7040059. R.R. would like to thank CNPq for partial financial
support.

\end{document}